\documentclass[%
 reprint,
superscriptaddress,
 amsmath,amssymb,
 aps,
]{revtex4-2}

\usepackage{graphicx}
\usepackage{dcolumn}
\usepackage{bm}
\usepackage{amsmath,amsfonts}
\usepackage{algorithmic}
\usepackage{algorithm}
\usepackage{array}
\usepackage[colorlinks=true, allcolors=blue]{hyperref}
\usepackage[caption=false,font=normalsize,labelfont=sf,textfont=sf]{subfig}
\usepackage{textcomp}
\usepackage{url}
\usepackage{verbatim}
\usepackage{graphicx,xcolor,ulem,todonotes}

\begin{document}

\preprint{APS/123-QED}

\title{In-Field Comparison between G.652 and G.655 Optical Fibers for Polarization-Based \\ Quantum Key Distribution}

\author{Costantino Agnesi}
    \email{costantino.agnesi@unipd.it}
    \affiliation{Dipartimento di Ingegneria dell'Informazione, Universit\`a degli Studi di Padova, Via Gradenigo 6B - 35131 Padua, Italy.}
\author{Massimo Giacomin}
    \affiliation{Dipartimento di Ingegneria dell'Informazione, Universit\`a degli Studi di Padova, Via Gradenigo 6B - 35131 Padua, Italy.}
\author{Daniele Sartorato}
    \affiliation{Dipartimento di Ingegneria dell'Informazione, Universit\`a degli Studi di Padova, Via Gradenigo 6B - 35131 Padua, Italy.}
    \affiliation{Telebit S.p.A.,  Via Marco Fanno 1 - 31030 Casier (TV), Italy}
\author{Silvia Artuso}
    \affiliation{Telebit S.p.A.,  Via Marco Fanno 1 - 31030 Casier (TV), Italy}
\author{Giuseppe Vallone}
    \affiliation{Dipartimento di Ingegneria dell'Informazione, Universit\`a degli Studi di Padova, Via Gradenigo 6B - 35131 Padua, Italy.}
\author{Paolo Villoresi}
    \affiliation{Dipartimento di Ingegneria dell'Informazione, Universit\`a degli Studi di Padova, Via Gradenigo 6B - 35131 Padua, Italy.}



\date{\today}

\begin{abstract}
Integration of Quantum Key Distribution (QKD) in existing telecommunication infrastructure is crucial for the widespread adoption of this quantum technology, which offers the distillation of unconditionally secure keys between users. In this letter, we report a field trial between the \textit{Points of Presence} (POPs) placed in Treviso and in Venezia - Mestre, Italy, exploiting the QuKy commercial polarization-based QKD platforms developed by ThinkQuantum srl and two different standards of single-mode optical fibers, i.e. \textit{G.652} and \textit{G.655}, as a quantum channel. In this field trial, several configurations were tested, including the co-existence of classical and quantum signals over the same fiber, providing a direct comparison between the performances of the G.652 and G.655 fiber standards for QKD applications.   \\
\end{abstract}

\maketitle


\section{Introduction}
Quantum Key Distribution (QKD) is a quantum communication protocol that allows users to distill a secret key with unconditional security~\cite{BB84}. 
Differently from computationally-secure classical algorithms used in our current encryption systems, QKD offers long-term privacy guaranteed by the laws of Quantum Mechanics, and is not threatened by algorithmic and technological advances for both classical and quantum computation~\cite{Pirandola2019rev}.
In fact, the strategic importance of QKD is highlighted by recent breakthroughs in quantum computing that accelerate the development of large-scale quantum computers capable of factorizing large prime numbers~\cite{Shor1997} that underlie our public cryptography schemes.
Furthermore, QKD is the first quantum communication protocol to have reached industrialization and commercialization. This has led to several national and international initiatives that have incentivize the deployment of QKD systems in our telecommunications networks.

Two of the most commonly exploited optical fibers that are widely implemented in telecommunications are those described by the standards ITU-T G.652~\cite{G652} and ITU-T G.655~\cite{G655} enforced by the International Telecommunication Union. 
Both fiber types exhibit single-mode operations for both the 1310 nm and 1550 nm bands, and represent good choices for long-haul links.
 The dispersion properties of these two technologies are the aspect in which they differ the most. G.652 fibers are optimized for use in the 1310 nm band since they have zero dispersion at that wavelength. On the contrary, G.655 fibers are more suitable for the employment in the 1550 nm wavelength region due to their reduced dispersion value in the C-band (1530-1660 nm). G.655 fibers have a higher refractive index leading to a larger numerical aperture and a wider acceptance angle, parameters that make these devices better suited for use in scenarios where it is challenging to control the launch conditions, such as in long distance or undersea communications. Furthermore, the G.655 show easy implementation with Erbium Doped Fiber Amplifiers (EDFA), making them appropriate for Wavelength Division Multiplexing (WDM) communication systems.

 \begin{figure}[htbp]
    \centering
    \includegraphics[scale=0.04]{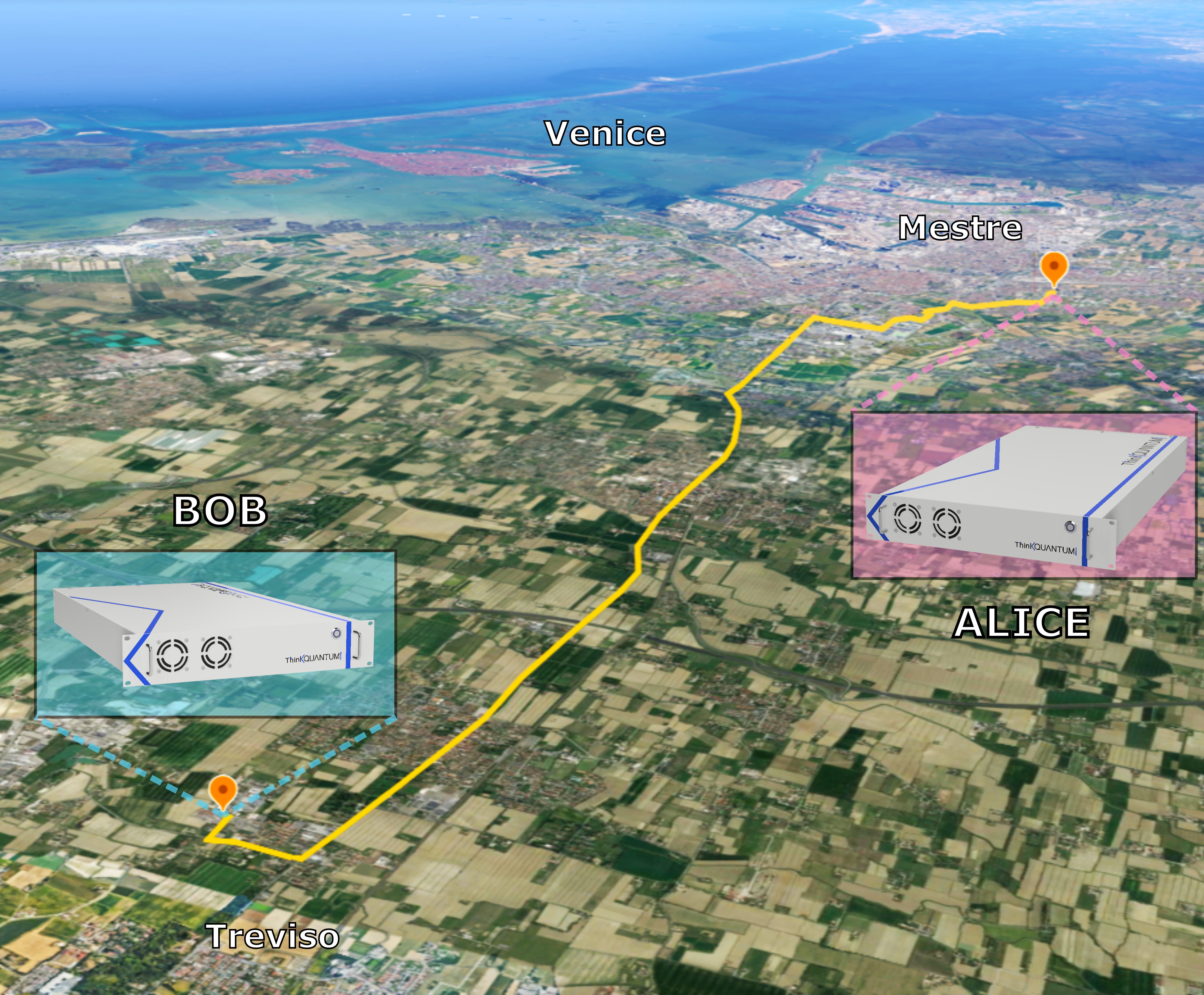}
    \caption{Aerial view of the optical network exploited for the reported results. The two G.655 and G.652 optical fibers cover a link of about $19$km between the cities of Treviso and Venice - Mestre, Italy. Map data: \copyright 2022 Google.}
    \label{fig:aerial_view}
\end{figure}

To the best of our knowledge, a direct comparison between the performance of G.652 and G.655 fibers for QKD applications with polarization encoding has yet to be performed. This study, however, is of interest to the QKD community, as both standards are widely employed in telecommunications networks.  In this letter, we report the results of a QKD field-trial exploiting both G.652 and G.655 deployed between two urban centers in the Veneto region of Italy. The QKD system used in the trial was the QuKy systems of \href{https://www.thinkquantum.com/}{ThinkQuantum srl}, which implement the BB84 protocol~\cite{BB84} exploiting polarization encoding. By the QuKy system we performed a 24-hour trial for both the G.652 and the G.655 fiber channels in a ``dark-fiber'' configuration. Additionally, a co-existance test between quantum and classical signals was performed in both fibers.
 
\section{Methodology}
\subsection{Optical network under test}
The optical network considered in this study is represented by a deployed optical channel owned by the private telecom provider Retelit S.p.A. and operated by Telebit S.p.A.,
that connects the \textit{Points of Presence} (POP) in the cities of Treviso and Venezia - Mestre for a distance of approximately $19$ km, as it can be observed in Fig.~\ref{fig:aerial_view}.\\ The two optical fibers are installed in parallel, and for this reason are considered to be exposed to the same macroscopic mechanical and thermal stresses. This configuration is optimal for this study, as it allows for a parallel comparison between the two systems, including channel perturbations due to the external environment. Lastly, these fibers are buried underground, making them less affected by optical and mechanical disturbances. From a usage point of view, the G.655 fiber used in the study is part of a 36 fiber bundle, composed of 3 micro-tubes with 94\% occupation of active data transmission. Instead, the G.652 fiber used in the study is part of a 72 fiber bundle, composed of 6 micro-tubes, with a much lower occupation rate of only 8\%.

\renewcommand{\arraystretch}{1.5} 
\begin{table}[htbp]
    \centering
    \begin{tabular}{|c||c|c|}
        \cline{2-3}
        \multicolumn{1}{c||}{} 
        & \multicolumn{2}{c|}{\bf Fiber}\\
        \cline{2-3}
        \multicolumn{1}{c||}{} 
        & \textbf{G.652} & \textbf{G.655}\\
        \hline
        \hline
        $\alpha$ [dB/km] @ 1310 nm & $0.3251\pm0.0001$ & $0.3814\pm0.0001$ \\
        \hline 
        $\alpha$ [dB/km] @ 1550 nm & $0.1862\pm0.0001$ & $0.2383\pm0.0001$ \\
        \hline
        $\alpha^{tot}$ [dB] @ 1310 nm & $6.0764\pm0.0005$ & $6.9420\pm0.0005$ \\
        \hline 
        $\alpha^{tot}$ [dB] @ 1550 nm & $3.4810\pm0.0005$ & $4.3380\pm0.0005$ \\
        \hline
        Total Length [km] & $18.692\pm0.001$ & $18.201\pm0.001$ \\
        \hline
    \end{tabular}
    \caption{Results of the OTDR analysis for both fibers.}
    \label{tab:otdr_results}
\end{table}
\renewcommand{\arraystretch}{1}

A preliminary characterization of the two optical fibers considered in this study  was performed in order to foresee their performances in classical and quantum communications. For this, an optical-time domain reflectometer (OTDR) test was firstly conducted, providing the results reported in Tab. \ref{tab:otdr_results}. As expected, both fibers manifest a lower attenuation coefficient $\alpha$ for  $1550$ nm wavelength, with respect to the $1310$ nm. Furthermore, according to the inferred results the standard G.652 offers less attenuation with respect to its counterpart, regardless of the considered wavelength.\\ The slight length differences observed in the two fibers can be explained by the diversity between the end reels placed at the end of each link during the OTDR analysis.

The second phase of the characterization of the optical network consisted in performing a \textit{Polarization Drift Measurement}, which allows to understand the stability of the specific considered optical fiber as a quantum channel in the implementation of a QKD protocol. 
As a matter of fact, quantum cryptographic techniques based on polarization states as encoding strategy require some polarization stability in order to retrieve the information shared among the users legitimately connected to the network. 
However, if the drift is slow enough, it is possible to characterize the rotation of polarization introduced by the birefringence of the fiber and restore the initial information sent through it using polarization controllers.\\
Polarization drift characterization was performed by injecting 1550 nm polarized light into both fibers and measuring the polarization state after the fiber propagation using a polarimeter. We estimated a drift of around $23$~mrads per second for both G.652 and G.655 fibers. This value was calculated as the scalar product computation between the subsequent samples obtained every second.   
The inferred result is well compatible with the polarization tracking capabilities of polarization-based QKD receivers~\cite{Ding2017}.

\begin{figure}[htbp]
    \centering
    \includegraphics[scale=0.65]{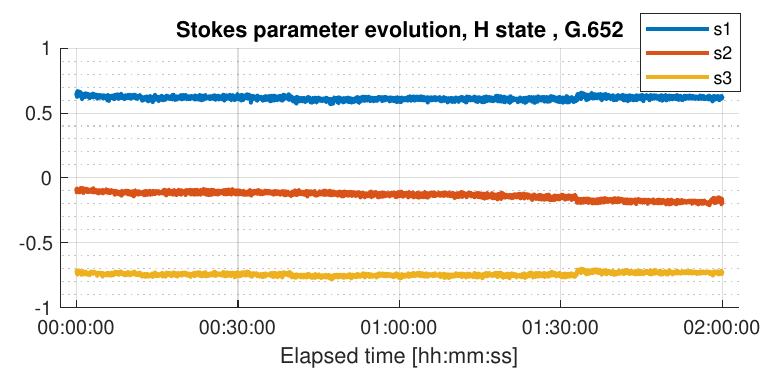}
    \includegraphics[scale=0.65]{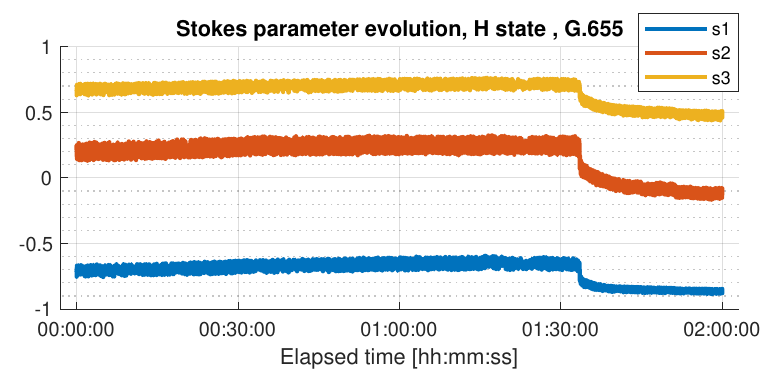}
    \caption{Evolution of the Stokes parameters in time for the two considered fiber standards.}
    \label{fig:stokes_drift}
\end{figure}

Furthermore, with these measurements we verified the robustness against polarization fluctuation of the G.652 and G.655, showing long-term stability without significant polarization rotation of the transmitted quantum states, as can be proven by \ref{fig:stokes_drift}. 

\subsection{Quantum Key Distribution system}
The QKD device used in this field trial was the QuKy system developed by ThinkQuantum srl, which implements the BB84 protocol encoded with three states and one decoy polarization~\cite{Grunenfelder2018}. This simplified protocol has the same level of security and performance as the original BB84 protocol~\cite{Tamaki2014}, whereas the one-decoy scheme can provide higher rates in the finite-key scenario compared to the two-decoy scheme~\cite{Rusca2018_APL}. The transmission of quantum signals is performed at 1550 nm, ensuring full compatibility with standard fiber-optic networks and components. The QuKy transmitter offers stable, low-error, and calibration-free polarization encoding by using the iPOGNAC modulator, which is described in Ref.~\cite{iPognac}. Furthermore, in contrast to proof-of-concept QKD demonstrations~\cite{FakeQKD}, the QuKy devices perform "real QKD" since phase randomization is performed via laser gain switching~\cite{PhysRevA.90.032320} 
and the randomness is provided by a source-device-independent Quantum Random Number Generator~\cite{Avesani2018} integrated inside the QKD transmitter (QRN2Qubit direct stream).

The quantum receiver decodes the quantum signals exploiting a polarimetric scheme that implements state projection into two mutually unbiased bases, whereas single-photon detection is performed using InGaAs/InP single-photon avalanche detectors (SPAD). Since both transmitter and receiver are fully contained in a 2U 19" Rackmount enclosure, the system can be easily integrated in telecom data centers, such as the Treviso and Venezia - Mestre POPs.
To compensate for the random and dynamic unitary transformation induced by the fiber-optical channel, the QuKy receiver integrates an all-fiber polarization controller. The correction signal applied to this polarization controller is given by a set of publicly disclosed qubits interleaved with the random qubit transmission. This allows for a quick determination of QBER and a rapid feedback cycle~\cite{Agnesi:20}, without requiring a dedicated feedback system composed of separate lasers and detectors. Furthermore, temporal synchronization is performed via the Qubits4Sync algorithm~\cite{PhysRevApplied.13.054041} to synchronize the frequency and absolute time between the transmitter and receiver device. This implies that the QuKy system \emph{does not} required any synchronization subsystem, which is usually implemented with a pulsed laser or GNSS clock to share an external time reference. 

\subsection{Field-trials configurations}
The fibers were tested in two configurations, labeled as \textit{dark-fiber} and \textit{CQ-coexistence}.
\vspace{0.2cm}

\subsubsection{Dark-Fiber}
This configuration represents an implementation of the QKD protocol in which the quantum channel and the classical channel are separated and realized with two different optical fibers. In this configuration, the classical signals were encoded at $1490$ nm using bidirectional SFP transceivers, while the quantum signals were emitted at $1550$ nm, as described in the previous section.

\subsubsection{CQ-coexistence} \label{sec:coexistane}
\begin{figure*}[htbp]
    \centering
    \includegraphics[scale=0.6]{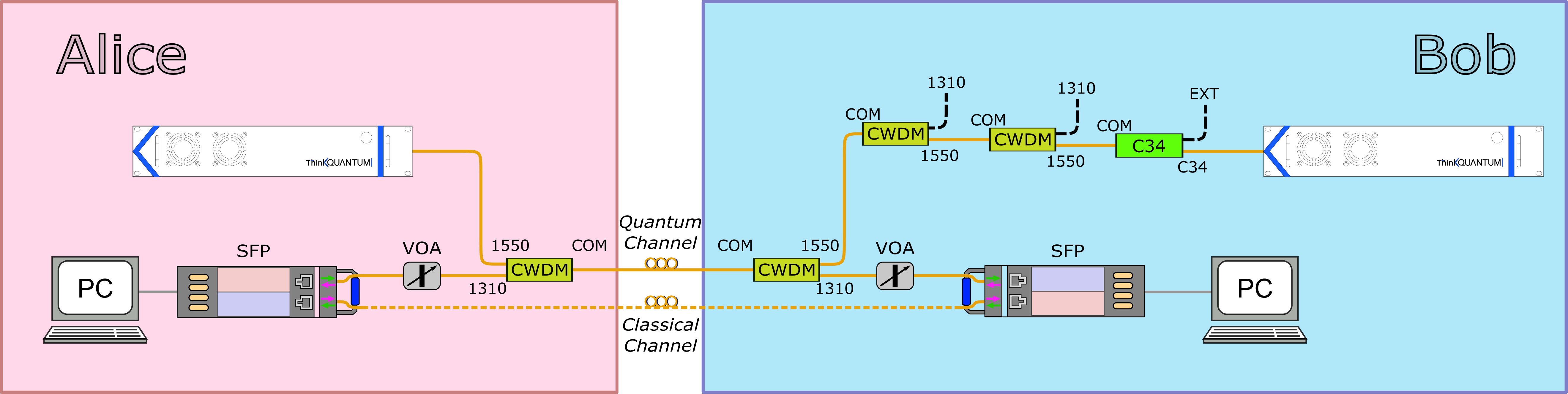}
    \caption{Graphical representation of the experimental setup required to analyze the optical fibers in Coexistence Duplex configuration. VOA: Variable Optical Attenuator, CWDM: Coarse Wavelength Division Multiplexer, C34: optical filter. The \textit{green arrows} in the SFP ports indicates the direction of the optical signal in the \textit{co-propagation} configuration, while the \textit{pink arrows} represent the \textit{counter-propagation} configuration.}
    \label{fig:coexistance}
\end{figure*}
In our QKD field-trial, the \textit{coexistence} configuration refers to the simultaneous operation of classical and quantum communications in the same fiber via wavelength division multiplexing (WDM). This scheme is of particular interest for the widespread adoption of QKD in telecommunication networks, since the integration of QKD with other communication systems can lead to a reduction in the high costs of the deployment and maintenance of quantum optical networks\cite{WDM}.

The setup used to implement the coexistence of QKD with classical communication is depicted in Fig.~\ref{fig:coexistance}. Classic communication was carried out using duplex SFP transceivers working at $1310$ nm. This wavelength was chosen to have sufficient spectral separation between the classical and quantum signals and to enable the use of commercial off-the-shelf WDM filters. In particular, coarse WDMs were used to merge and separate the $1310$ nm light from the $1550$ nm quantum signal, and a more dense WDM allowed for narrow (0.8 nm) filtering at the QKD wavelength (C34 according to the ITU G.694.1 spectral grid). Additionally, the duplex configuration of the SFP transceivers allowed us to test two different possible implementations. The first one was the co-propagation scheme where the quantum signals propagated together with the classical signal sent from Alice to Bob, while the classical signals sent from Bob to Alice exploited a separate optical fiber. The second implementation was the counterpropagation scheme, where the quantum signals, still sent from Alice to Bob, propagated in the same channel but in opposite direction of the classical signal sent from Bob to Alice, while the classical signals sent from Alice to Bob exploited a separate optical fiber.\\
In order to reduce the interference between the quantum and classical signals, a Variable Optical Attenuator (VOA) was inserted after the SFP responsible for emitting the light co-propagating with the QKD signal. This VOA was set at the maximum attenuation possible that still guaranteed the nominal 10 Gbps connection, which led to a launch power of the classical signal of approximately $-24$ dBm in the case of the G.652 fiber and about $-23$ dBm for the G.655 fiber.


\section{Results}
\subsection{QKD 24-hour test in dark fiber configuration}
\begin{figure}[htbp]
    \centering
    \includegraphics[scale=0.6]{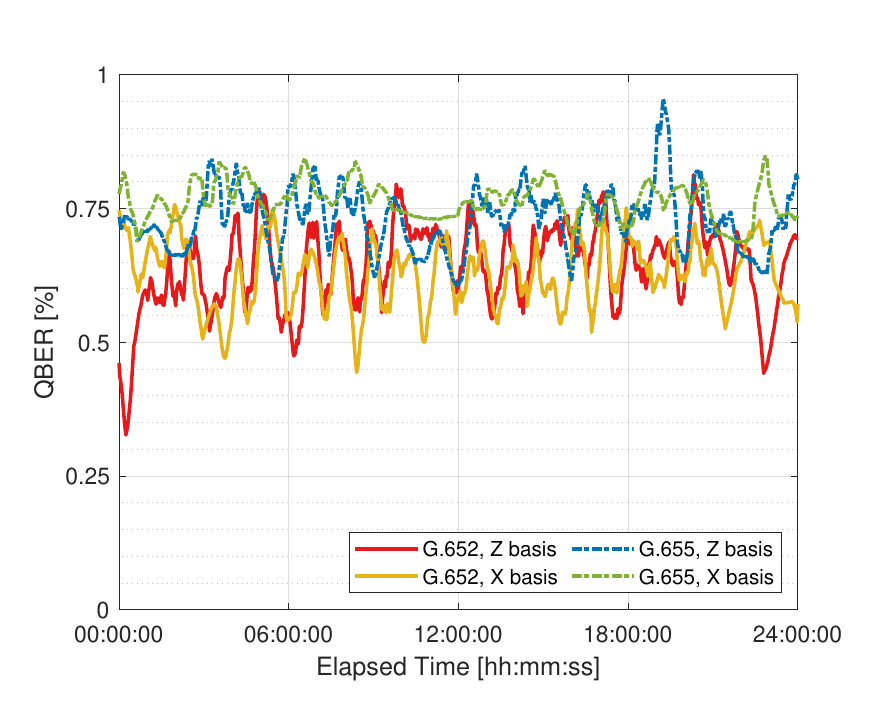}
    \caption{24 hours time evolution of the QBER considering the two \text{key} ($\mathcal{Z}$) and \textit{control} ($\mathcal{X}$) bases together with their lowerbounds.}
    \label{fig:qber}
\end{figure}
Between Wednesday, November 30\textit{th} 2022, and Thursday, December the 1\textit{st} 2022, we performed the first QKD test exploiting the G.652 as quantum channel, beginning at $12:00$ am (CET) and lasting for exactly 24 hours. Subsequently, at $01:00$ pm (CET) of Thursday, December the 1\textit{st}, the test on the G.655 fiber started, lasting until $01:00$ pm (CET) of Friday, December the 2\textit{nd}. The days of the test can be considered typical weekdays, all of which exhibit temperate winter weather with a daily average temperature of $7^\circ$.

\begin{figure}[htbp]
    \centering
    \includegraphics[scale=0.6]{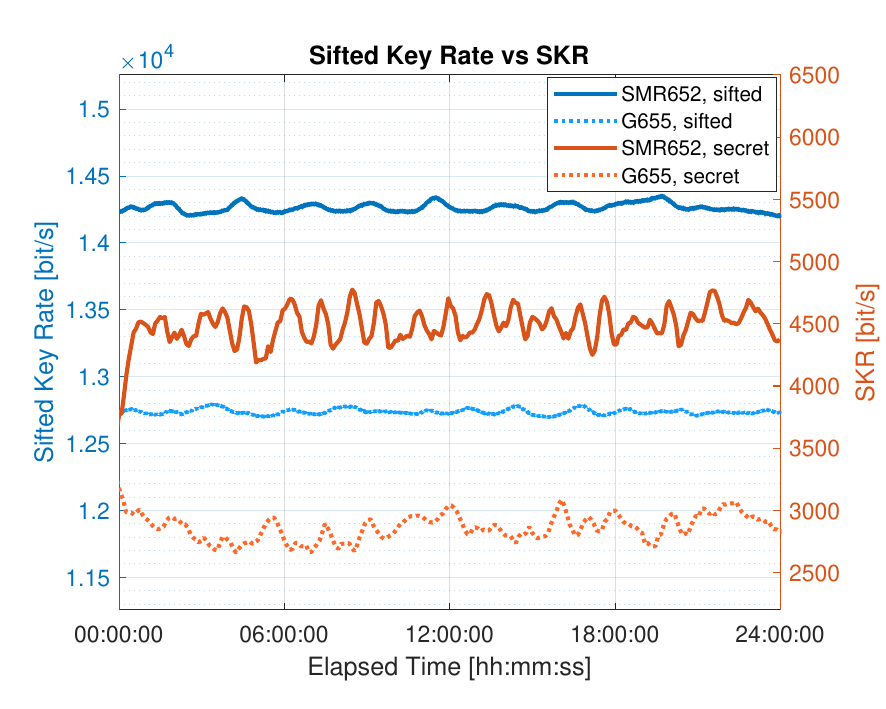}
    \caption{Time evolution of the \textit{Key Rates}: \textit{Sifted} in \textit{blue} on the left axis and \textit{Secret} in \textit{orange} on the right axis.}
    \label{fig:keyRates}
\end{figure}

To assess the performance of the QKD system, we monitored the main key performance indicators, i.e., \textit{Quantum Bit Error Rate} (QBER), \textit{Sifted Key Rate} and \textit{Secure Key Rate} (SKR).  The results on QBER are shown in Fig.~\ref{fig:qber}, while the Sifted Key Rate and SKR are shown in Fig.~\ref{fig:keyRates}, computed considering the finite-key effects with security parameter $\epsilon_{\rm sec}=10^{-15}$.


\renewcommand{\arraystretch}{1.5} 
\begin{table*}[htbp]
    \centering
    \begin{tabular}{|c||c|c|c|c|}
        \hline
        \quad \textit{Configuration} \quad & \quad \textrm{Co-P. G.652}\quad \quad & \textrm{Counter-P. G.652} \quad & \quad \textrm{Co-P. G.655} \quad \quad & \textrm{Counter-P. G.655} \quad \\[0.5ex]
        \hline
        \hline
        \textbf{QBER Z [$\%$]} & $1.7\pm1.4$ & $1.4\pm0.4$ & $2.9\pm0.7$ & $2.5\pm0.5$ \\
        \hline 
        \textbf{QBER X [$\%$]} & $1.9\pm0.8$ & $2.0\pm0.3$ & $3.7\pm1.1$ & $3.1\pm0.3$ \\
         \hline
        \textbf{SNR} & $40.2\pm2.4$ & $37.2\pm2.2$ & $17.9\pm1.2$ & $20.2\pm1.1$ \\
        \hline
        \textbf{Noise [$\mathrm{events}/s$]}& $7600\pm 600$ & $8200 \pm 400$ & $12000 \pm 1700$ & $11000 \pm 500$ \\
        \hline
        \textbf{Sifted KR} \textrm{ [kb/s]} & $10.8\pm1.1$ & $10.7\pm1.0$  & $7.8\pm0.4$  & $7.9\pm0.4$ \\
        \hline
        \textbf{Secure KR} \textrm{ [b/s]} & $2400 \pm 300$ & $2000 \pm 600$ & $21 \pm 15$ & $48 \pm 62$ \\
        \hline
    \end{tabular}
    \caption{Results of the coexistence test for both the co-propagation (\textit{Co-P.}) and counter-propagation (\textit{Counter-P.}) configurations, studied considering the two optical fibers under test, i.e. G.652 and G.655.}
    \label{tab:coexistence_results}
\end{table*}
\renewcommand{\arraystretch}{1}

Regarding the G.652 fiber we report an average QBER of $(0.64\pm0.30)\%$ in the key-generating $\mathcal{Z}$ basis and $(0.65~\pm~0.48)\%$ in the control state of the $\mathcal{X}$ basis. An average of $14264 \pm 1260$ of bits were sifted per second, resulting in an average secret key generation of $4460 \pm 357$ bits per second and a total generation of $392$ megabits of secret key material.
Instead, for the G.655 fiber, we report an average QBER of $(0.73 \pm 0.31)\%$ in the key-generating $\mathcal{Z}$ basis and $(0.78 \pm 0.19)\%$ in the control state. An average of $12736 \pm 1015$ bits per second were sifted resulting in an average secret key generation of $2869 \pm 236$ bits per second and a total generation of $296$ megabits of secret key material.\\
Small fluctuations in the QBER are interpreted as an expected effect of the birefringence of the optical fibers, which may be affected by predictable environmental stresses. These determine mild oscillations in the finale Sifted ad Secret Key Rates. However, the resulting QBER shows the robustness and suitability of the fibers under test, remaining below the threshold of $1\%$

\subsection{Co-existance Results}
On the afternoon of December, the 2\textit{nd}, after the completion of the 24 hour run in the G655 fiber, we performed a series of experiments to test the co-existance of QKD with classical communication on the same fiber. As thoroughly explained in Sect.~\ref{sec:coexistane}, these included co-propagation and counter-propagation of a 1310 nm classical signal together with the 1550 nm QKD signal.
The results of the coexistence experiment are reported in Tab.~\ref{tab:coexistence_results} where the most significant parameters as the QBERs, the SNR, the Sifted Key Rate and the
Secret Key Rate are shown for the two optical fibers under test. As before, the SKR was calculated using the finite-key approach.

\section{Discussion}


The Sifted and Secret Key Rate results reported for the dark fiber configuration are in line with expectations since the performance distinctions can be mainly explained by the difference in the total losses for both fiber standards. However, more considerations must be made to interpret the lower QBER observed in the G.652 fiber channel, which is due to a higher Signal-to-Noise Ratio (SNR) of $208.44 \pm 22.62$ compared to $105.19 \pm 8.79$ of the G.655 fiber. In fact, this last result is partially unexpected if just a total loss dependence on the SNR is taken into account. 
Indeed, the halving of the final SNR moving from G.652 to G.655 suggests further causes in addition to the 1 dB signal losses introduced by the NZD standard. 
An explanation could be based on the much higher occupation rate of the G.655 fiber bundle used in the experiment, which could have introduced crosstalk between adjacent fibers, as suggested by the increase of the observed noise level. However, we would like to stress that the overall QKD performance in both fiber standards can be thought to be excellent, especially when considering that they were obtained from deployed fibers used for day-to-day telecommunications between the Venezia - Mestre and Treviso POPs.  

The co-existence configurations showed a relevant distinction in performance between the G.652 and the G.655 standards. Whereas secure key generation was possible in all tested configurations for both fiber standards, the SKR obtained with the G.652 fiber was around two orders of magnitude higher than with G.655 standard. This difference in SKR can be attributed to a higher QBER observed in the G.655, also reflected in a higher SNR for the G.652 fiber. This difference in SNR is only partially explained by the lower signal rate observed in the G.655 due to its higher attenuation. Higher noise levels were also observed in the G.655 channel, mainly due to a higher stimulated Raman scattering~\cite{RamanScatteringBook} caused by the higher launch power of the classical channel.   


\section{Conclusions}
In summary, we have assessed the performances of deployed G.652 and G.655 fibers used as quantum channel for a QKD protocol based on polarization encoding implemented by ThinkQuantum's QuKy platform. This study constitutes a first attempt to classify and characterize different standards in optical fibers for quantum communications, hence further theoretical and experimental studies must be performed to understand the difference in terms of noise and transmissivity when alternative physical links are implemented.

Furthermore, additional optimizations should be performed for the coexistance trial, especially when exploiting the more lossy G.655 standard. As a matter of fact, in the case of exploiting more than one wavelength, several non-linear phenomena may occur, such as Raman Scattering or Four-Wave Mixing, that collectively contribute to increase the noise at the detection side.

Lastly, this work represents a concrete example of a fully functioning commercial QKD system being integrated in a deployed and operating network, demonstrating that QKD technology can be exploited in real-case scenarios.

\section*{Acknowledgments}
We would like to thank \href{https://www.telebit.it/}{Telebit S.p.A.} for technical support and general assistance in the measurement campaign, with a particular effort for the OTDR metering, as well as \href{https://www.retelit.it/it/home}{Retelit S.p.A.} for having provided the characterized optical network.
\vspace{0.3cm}

\pagebreak
\textbf{Funding}:
PhD scholarship of MG was co-funded by \textit{Telebit S.p.A.} and MIUR - Ministerial Decree 352/2022. This work was partially funded by the European Union (QUID project, GA 101091408). Views and opinions expressed are however those of the author(s) only and do not necessarily reflect those of the European Union or the European Commission-EU. Neither the European Union nor the granting authority can be held responsible for them.


\bibliography{bibliography}

\end{document}